\documentclass[a4paper,11pt]{article}
\usepackage{pos}
\usepackage{hyperref}
\usepackage{lineno}

\title{Astro-COLIBRI: An Advanced Platform for Real-Time Multi-Messenger Astrophysics}
 \ShortTitle{Astro-COLIBRI}

\author*[a]{Fabian Sch\"ussler}
\author[a]{M. de Bony de Lavergne}
\author[a,b]{A. Kaan Alkan}
\author[a]{J. Mourier}
\author[c]{P. Reichherzer}

\affiliation[a]{IRFU, CEA, Université Paris-Saclay, Gif-sur-Yvette, France}
\affiliation[b]{Laboratoire Interdisciplinaire des Sciences du Numérique, CNRS, Université Paris-Saclay, 91405 Orsay, France}
\affiliation[c]{Department of Physics, University of Oxford, Oxford OX1 3PU, United Kingdom}

\emailAdd{fabian.schussler@cea.fr}

\abstract{Observations of transient phenomena like Gamma-Ray Bursts (GRBs), Fast Radio Bursts (FRBs), stellar flares and explosions (novae and supernovae), combined with the detection of novel cosmic messengers like high-energy neutrinos and gravitational waves has revolutionized astrophysics over the last years. The discovery potential of both ulti-messenger and multi-wavelength follow-up observations as well as serendipitous observations could be maximized with a novel tool which allows for quickly acquiring an overview over relevant information associated with each new detection. Here we present Astro-COLIBRI, a novel and comprehensive platform for this challenge.

Astro-COLIBRI's architecture comprises a public RESTful API, real-time databases, a cloud-based alert system and a website as well as apps for iOS and Android as clients for users. Astro-COLIBRI evaluates incoming messages of astronomical observations from all available alert streams in real time, filters them by user specified criteria and puts them into their MWL and MM context. The clients provide a graphical representation with an easy to grasp summary of the relevant data to allow for the fast identification of interesting phenomena, provides an assessment of observing conditions at a large selection of observatories around the world, and much more.

Here the key features of Astro-COLIBRI are presented. We outline the architecture, summarize the used data resources, and provide examples for applications and use cases. Focussing on the high-energy domain, we'll discuss the use of the platform in searches for high-energy gamma-ray counterparts to high-energy neutrinos, gamma-ray bursts and gravitational waves.
}

\ConferenceLogo{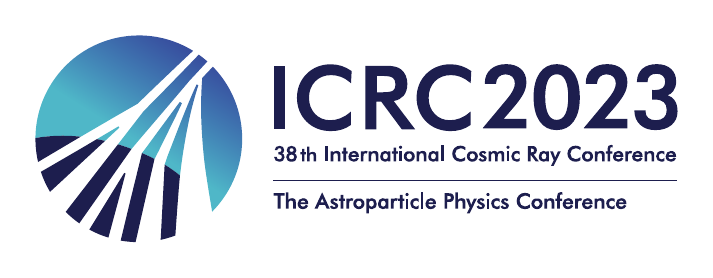}

\FullConference{%
38th International Cosmic Ray Conference (ICRC2023)\\
  26 July - 3 August, 2023\\
  Nagoya, Japan}

\begin{document}
\maketitle

\begin{figure}[t!]
\begin{center}
\includegraphics[width= 0.75\textwidth]{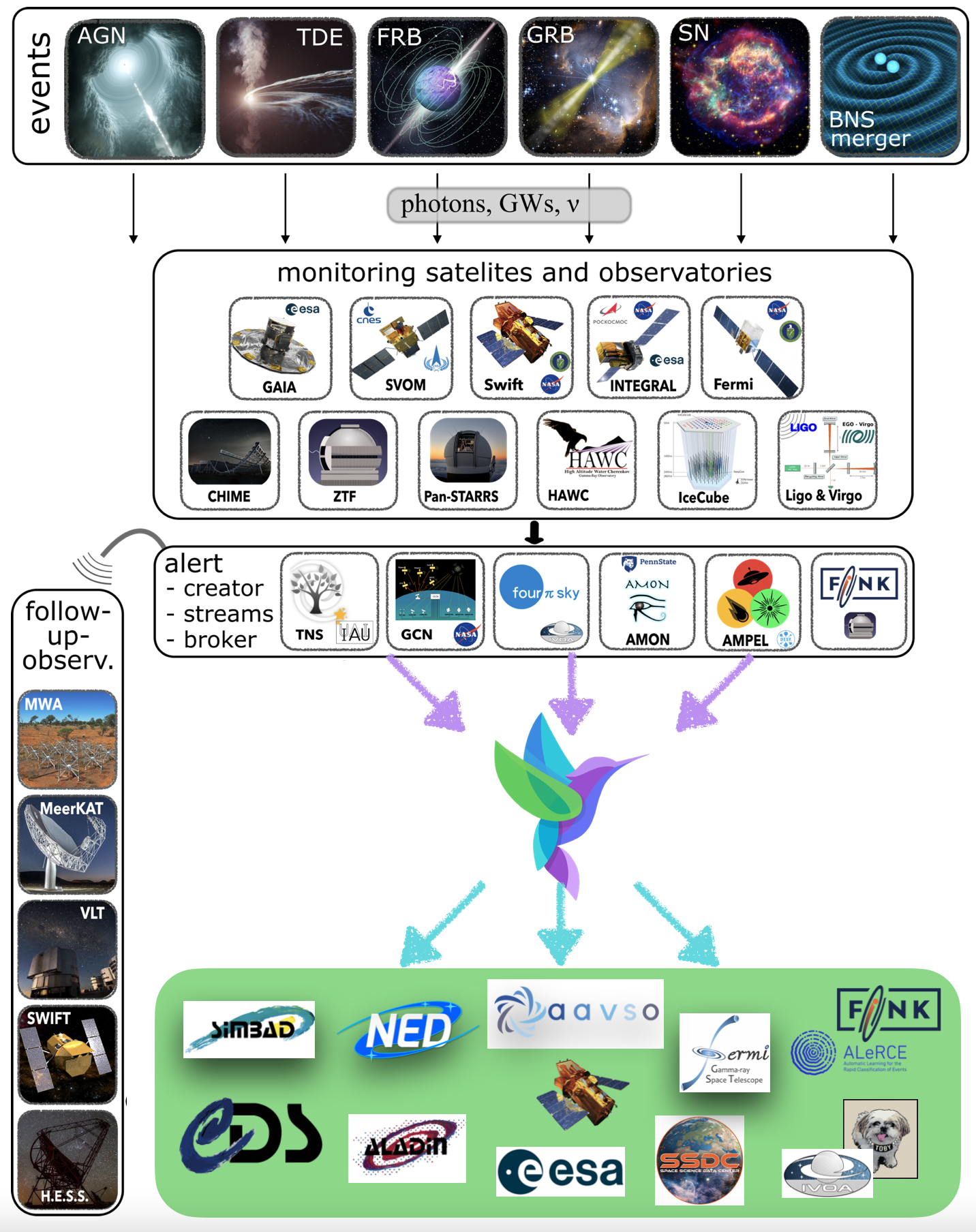}
\caption{Astro-COLIBRI is conveived as top-level platfrom fully integrated in the global multi-wavelength and multi-messenger landscape.}
\label{fig:landscape}
\end{center}
\end{figure}

\section{Introduction}

Astro-COLIBRI is an innovative platform designed to facilitate the study of transient astrophysical events by integrating real-time multi-messenger observation tools in a comprehensive and user-friendly graphical interface. By bundling and evaluating alerts about transient events from various channels, Astro-COLIBRI streamlines the process of coordinating follow-up observations, enabling professional and amateur astronomers alike to better understand the nature of these events through complementary observational data. Astro-COLIBRI supports a wide range of astrophysical source classes, including Active Galactic Nuclei (AGN), Gamma-ray Bursts (GRBs), Fast Radio Bursts (FRBs), Gravitational Waves (GWs), High-energy Neutrinos, Optical Transients (OT), Supernovae (SN), and more. By incorporating multi-messenger services, catalogs, and alerts from various observatories and systems, the platform facilitates seamless collaboration and data sharing among the astronomy community.

Astro-COLIBRI fully integrates into the global landscape of multi-wavelength and multi-messenger studies of transient phenomena. A top level view is given in Fig.~\ref{fig:landscape}. The platform is linked to a large variety of alert streams and brokers distributing information about new detections and classifications of transient phenomena. It summarizes this informations and puts it in the context of other time domain and archival data. Each event is also directly linked to additional, external services, information sources and expert platforms. Dedicated tools allow for efficient preparations of follow-up observations with a large variety of observatories. 

We here present an overview of the main features of the Astro-COLIBRI platform, focussing on recent additions and use-cases that exceed the description given in~\citep{2021ApJS..256....5R, 2023Galax..11...22R}.

\section{Main features}
The Astro-COLIBRI platfrom can be accessed through two main channels: programmatically through public API endpoints and via graphical user interfaces. 

\subsection{API endpoints}
The central backend of the Astro-COLIBRI plattform is a cloud-based RESTful API service. It is publically accessible at \url{https://astro-colibri.science}. The site acts also as central documentation hub for both the API (cf. \href{https://astro-colibri.science/apidoc}{API Doc} subpage) and the user interfaces (cf. \href{https://astro-colibri.science/documentation}{Documenation} subpage). The page also provides links to additional ressources like tutorial and onboarding videos, social media links, merchandise, etc. 

The various API endpoints allow to integrate the information collected by Astro-COLIBRI into external environments and tools. Use cases range from dedicated tools made available to burst advocates and shift-crews at large, professional observatories to observation planning tools for amateur astronomers.

\begin{figure}[t!]
    \centering
    \begin{minipage}{0.62\textwidth}
        \centering
        \includegraphics[width=0.9\textwidth]{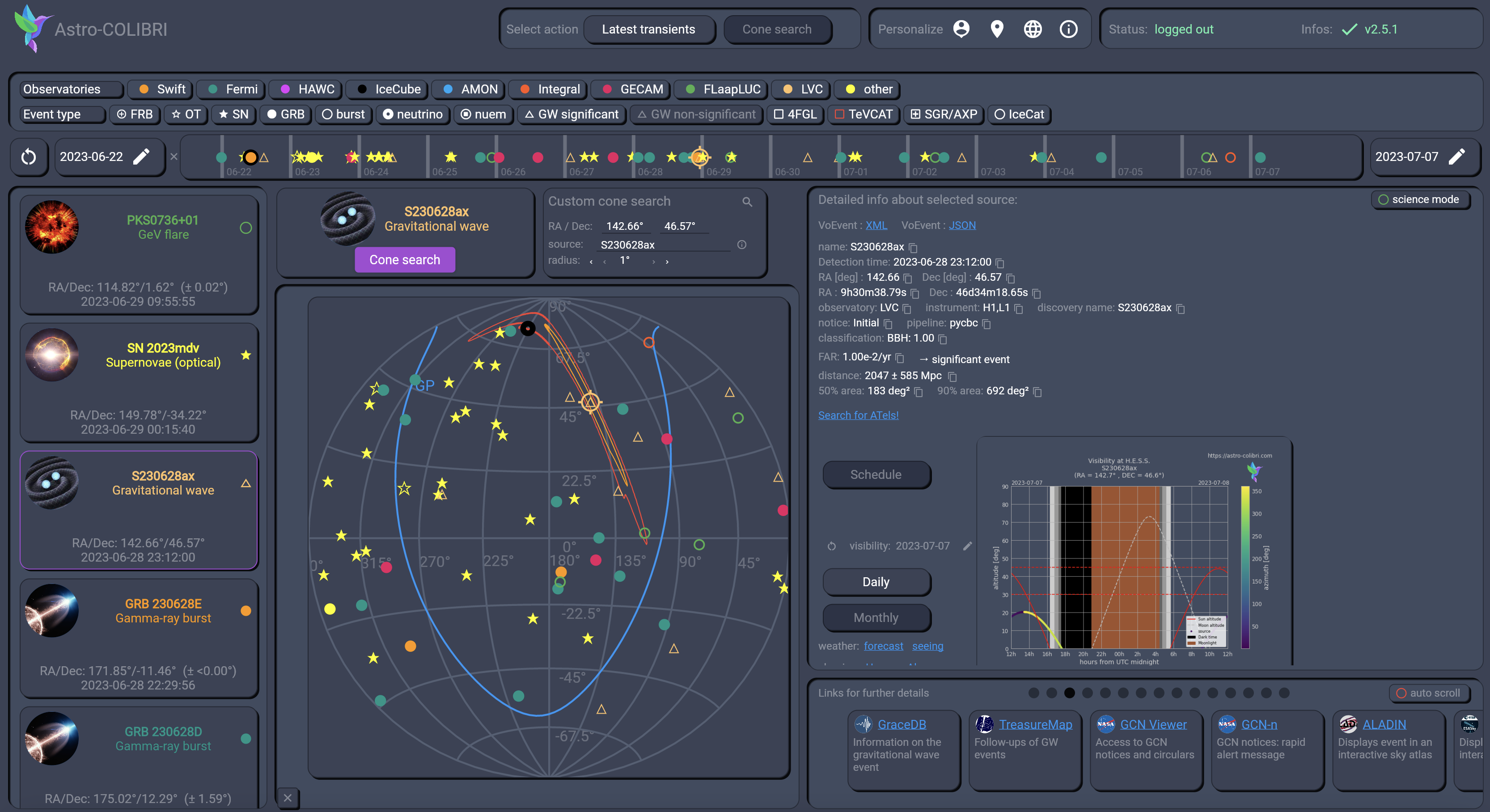} 
        \caption{Web frontend}\label{fig:web}
    \end{minipage}\hfill
    \begin{minipage}{0.36\textwidth}
        \centering
        \includegraphics[width=\textwidth]{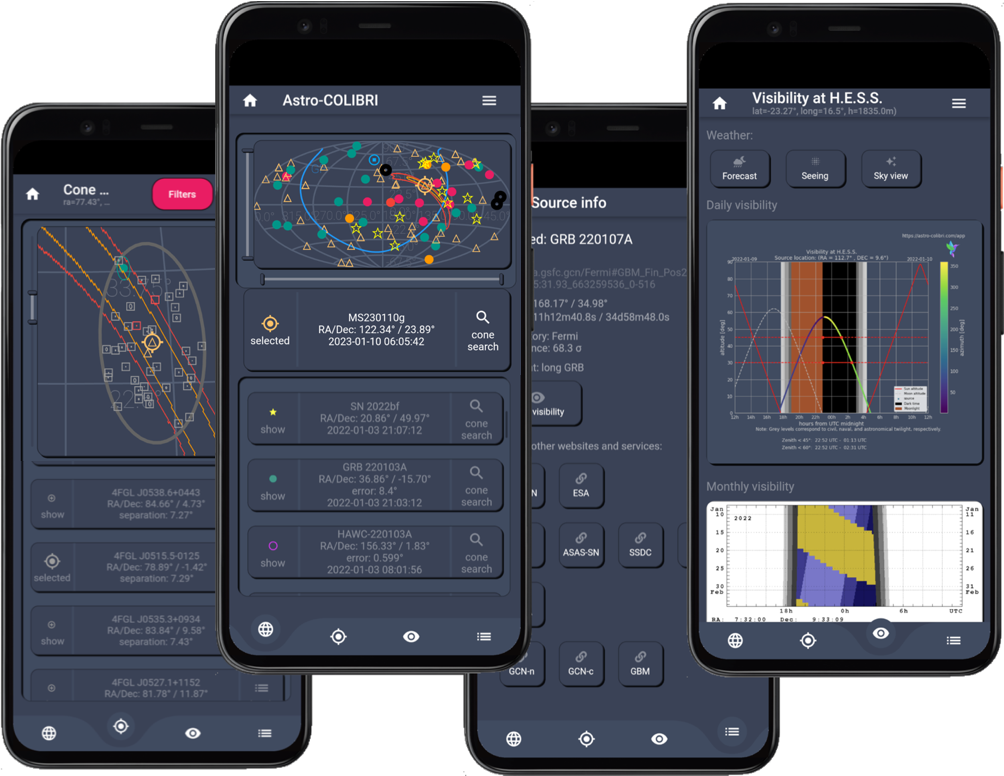} 
        \caption{Smartphone app}\label{fig:mobile}
    \end{minipage}
\end{figure}

\subsection{Graphical interfaces}
Astro-COLIBRI aims to provide an easy to use graphical experience for both, professional and amateur astronomers. The interface is available on the web (\url{https://astro-colibri.com}) and as application on both major mobile platforms, \href{https://play.google.com/store/apps/details?id=science.astro.colibri}{Android} and \href{https://apps.apple.com/us/app/astro-colibri/id1576668763}{iOS/iPadOS}. 

The interfaces can be customized with detailed settings and filter options. The~timeline (displayed in the top region of the web interface, cf. Fig.~\ref{fig:web}) contains all transient events in the user-defined time range that satisfy the user-set filters. These transients are also shown in the list on the left side and in the central sky map. The right info area displays further information about the selected event/source and lists direct links to other websites, light curves, event displays, energy spectra, etc. The website also allows to switch to a scientific mode using the {\it Science} button in the top right corner of the event information area. This activates the display of a short and long-term observability assessment for the specified follow-up observatory. 

\section{Use case examples}
The Astro-COLIBRI platform has been widely adopted across the professional and amateur astronomy communities. It is being used by burst advocates and astronomers on duty of large observatories around the world and has already enabled and facilitated multiple large observation campaigns, but also theoretical studies and citizen science projects. Here we give a very short snapshot of the most recent developments. Further examples are given on this page: \url{https://astro-colibri.science/usecases}.

\begin{figure}[t!]
    \centering
     \begin{minipage}{0.4\textwidth}
        \centering
        \includegraphics[width=\textwidth]{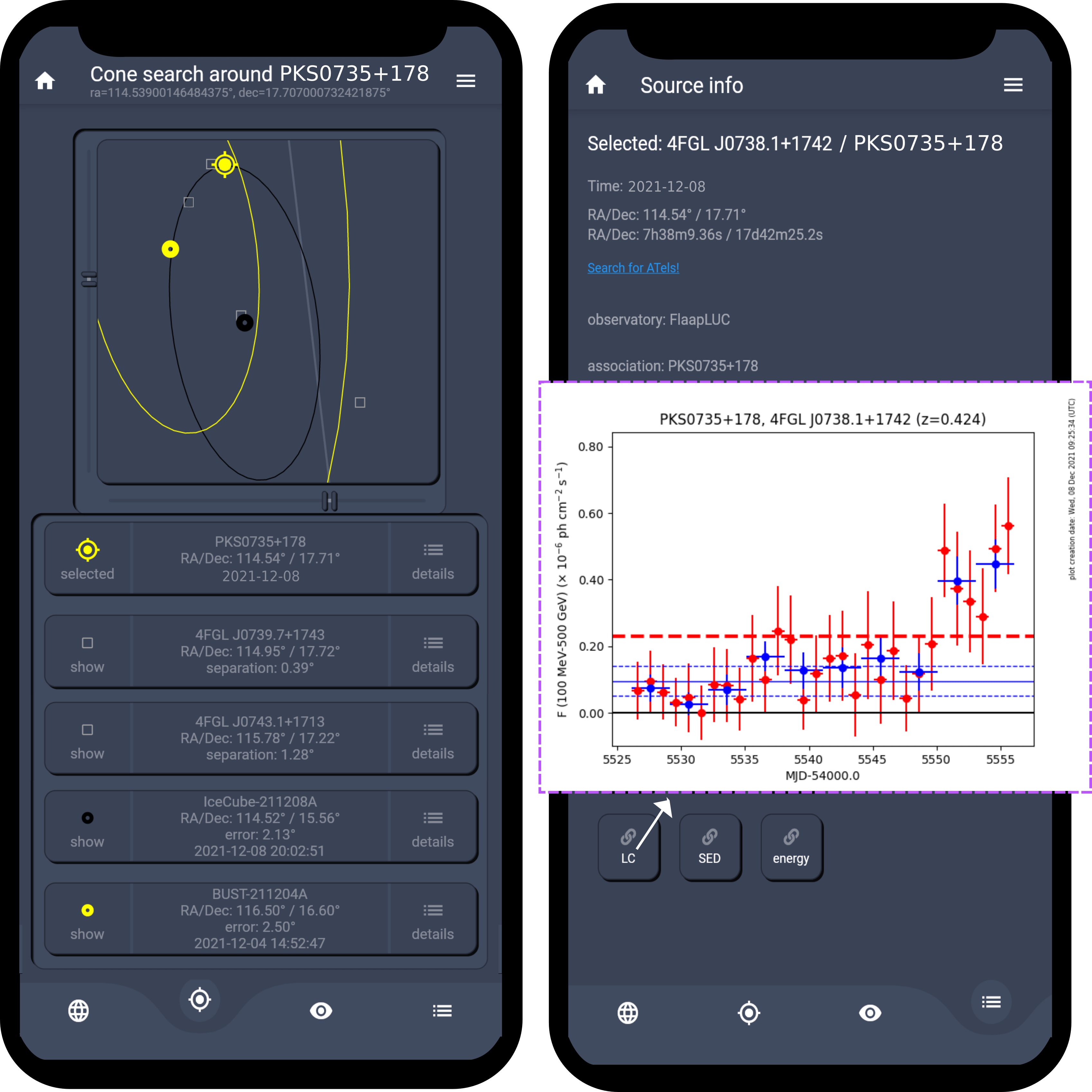} 
        \caption{Left: the Astro-COLIBRI view of PKS 0735+178 and the uncertainty region
 of IceCube-211208A (black ellipse). Right: the GeV lightcurve provided by FLaapLUC at the time of the neutrino detection.}\label{fig:pks_ac}
    \end{minipage} \hfill
    \begin{minipage}{0.55\textwidth}
        \centering
        \includegraphics[width=0.9\textwidth]{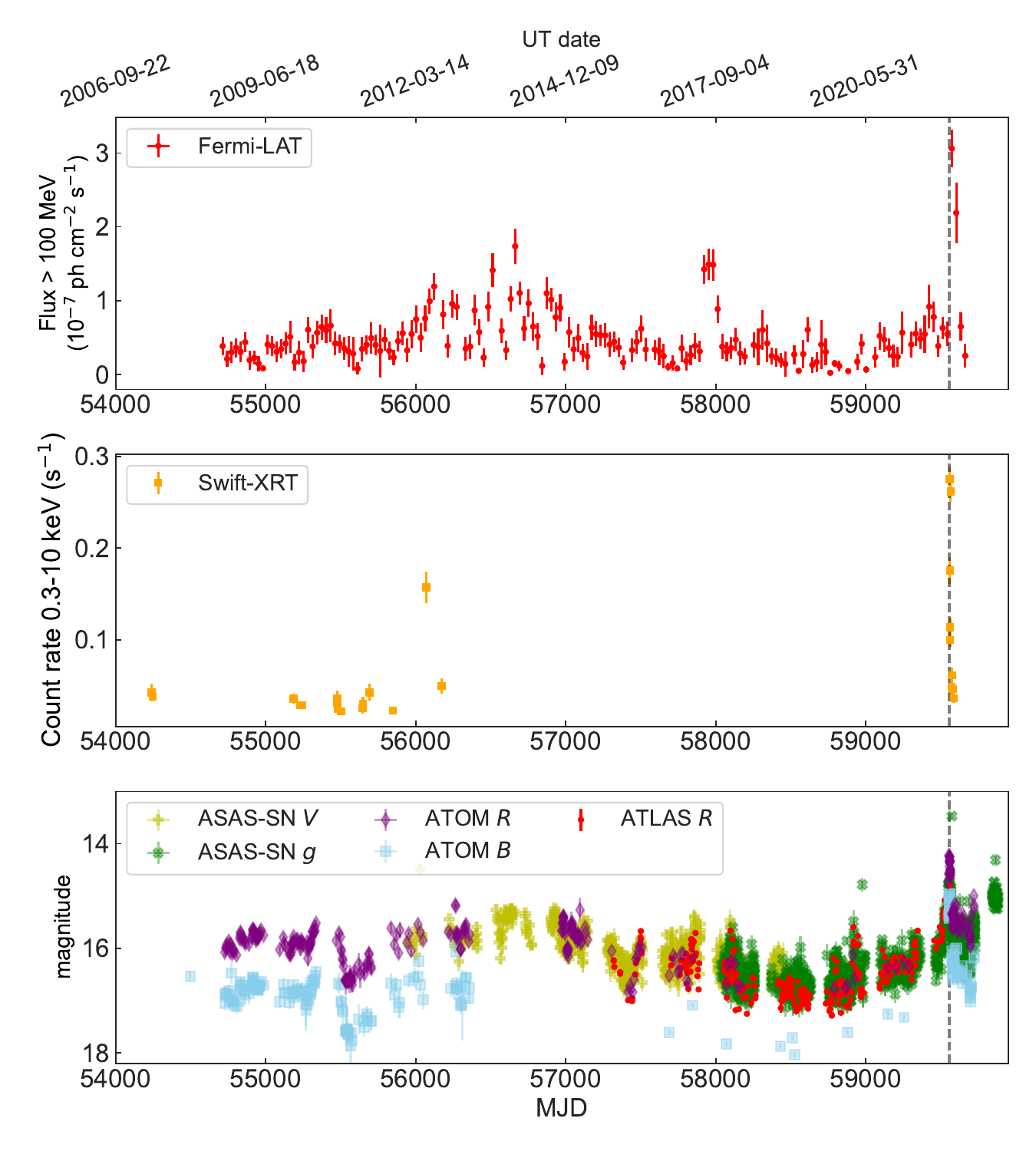} 
        \caption{MWL Lightcurve of PKS~0735+178 and the detection of the neutrino event IceCube-211208A (dashed line). From~\citep{2023arXiv230617819A}.}\label{fig:pks_lc}
    \end{minipage}
\end{figure}

\subsection{Searches for gamma-ray emission associated to high-energy neutrinos}
Multi-wavelenght observations are a promising way to localize the sources of high-energy neutrinos and thus pin-point the long-sought accelerators of high and ultra-high energy cosmic rays. A first glimpse for these studies, and starting point of the Astro-COLIBRI development, was the detection of a flaring blazar TXS~0506+056 in coincidence with the high-energy neutrino IceCube-170922A~\citep{IceCube2018MMA}.

While it took more than a week after the detection of IceCube-170922A to establish the connection with the flaring TXS~0506+056 and thus trigger an extensive multi-wavelenght campaign, with Astro-COLIBRI, it is now possible to check these correlations within seconds on any mobile device. The power of these tools is illustrated by the multi-messenger events that happened in December 2021: On December 8, 2021, the neutrino event IceCube-211208A was detected by the IceCube observatory. This event had an energy of 171 TeV and a 50.2\% probability of being of astrophysical origin. The event was announced in real-time via the Astro-COLIBRI notifications. Only a few hours earlier, another notification (via the automatic analysis of data recorded by Fermi-LAT provided by the FLaapLUC~\citep{Lenain:2018} pipeline) had announced the beginning of a gamma-ray flare of the blazar PKS 0735+178, located just outside the 90\% error region of the neutrino event. Both events were approximately four hours later complemented by the Baikal-GVD experiment which detected a high-energy neutrino candidate event with an energy of 43 TeV. Furthermore, the KM3NeT neutrino detectors found a candidate up-going muon neutrino (18 TeV) on December 15, spatially coinciding with PKS 0735+178 with a p-value of 0.14. The Baksan Underground Scintillation Telescope reported the observation of a GeV neutrino candidate event prior to IceCube-211208A.

The cone search view (shown in Fig.~\ref{fig:pks_ac}) of Astro-COLIBRI allowed to rapdily establish these coincidences and thus trigger dedicated follow-up observations across the electromagnetic spectrum. These observations have revealed activity in the radio, optical, X-ray, and GeV gamma-ray bands, potentially associated with the neutrino event IceCube-211208A. Target-of-opportunity observations with imaging atmospheric Cherenkov telescopes, namely VERITAS and H.E.S.S., yielded stringent upper limits above 100 GeV. Detailed results of these observations that contribute to the study of the blazar PKS 0735+178 and its potential connection to the neutrino event IceCube-211208A are given in~\citep{2023arXiv230617819A}.

\subsection{Follow-up observations of gravitational waves}
Gravitational waves (GWs) and the associated multi-wavelength emissions allow unprecedented studies but are challenging due to the large localisation uncertainty regions provided by the gravitational wave interferometers. Follow-up observations aiming to localise the source of the GW emission thus require dedicated and optimized scheduling procedures. One such scheduling tool is {\it tilepy}, a packaged that has been developed within the H.E.S.S. collaboration~\citep{2021JCAP...03..045A}. The code is publicly available (cf. \href{https://github.com/astro-transients/tilepy}{tilepy@GitHub}). In addition, Astro-COLIBRI is providing an open, cloud-based and ready-to-use installation at \url{https://tilepy.com}. The scheduling calculation can be requested easily via an API endpoint. Further details about {\it tilepy} are given in~\citep{tilepy_icrc2023}. 

\begin{figure}[t!]
    \centering
     \begin{minipage}{0.3\textwidth}
        \centering
        \includegraphics[width=\textwidth]{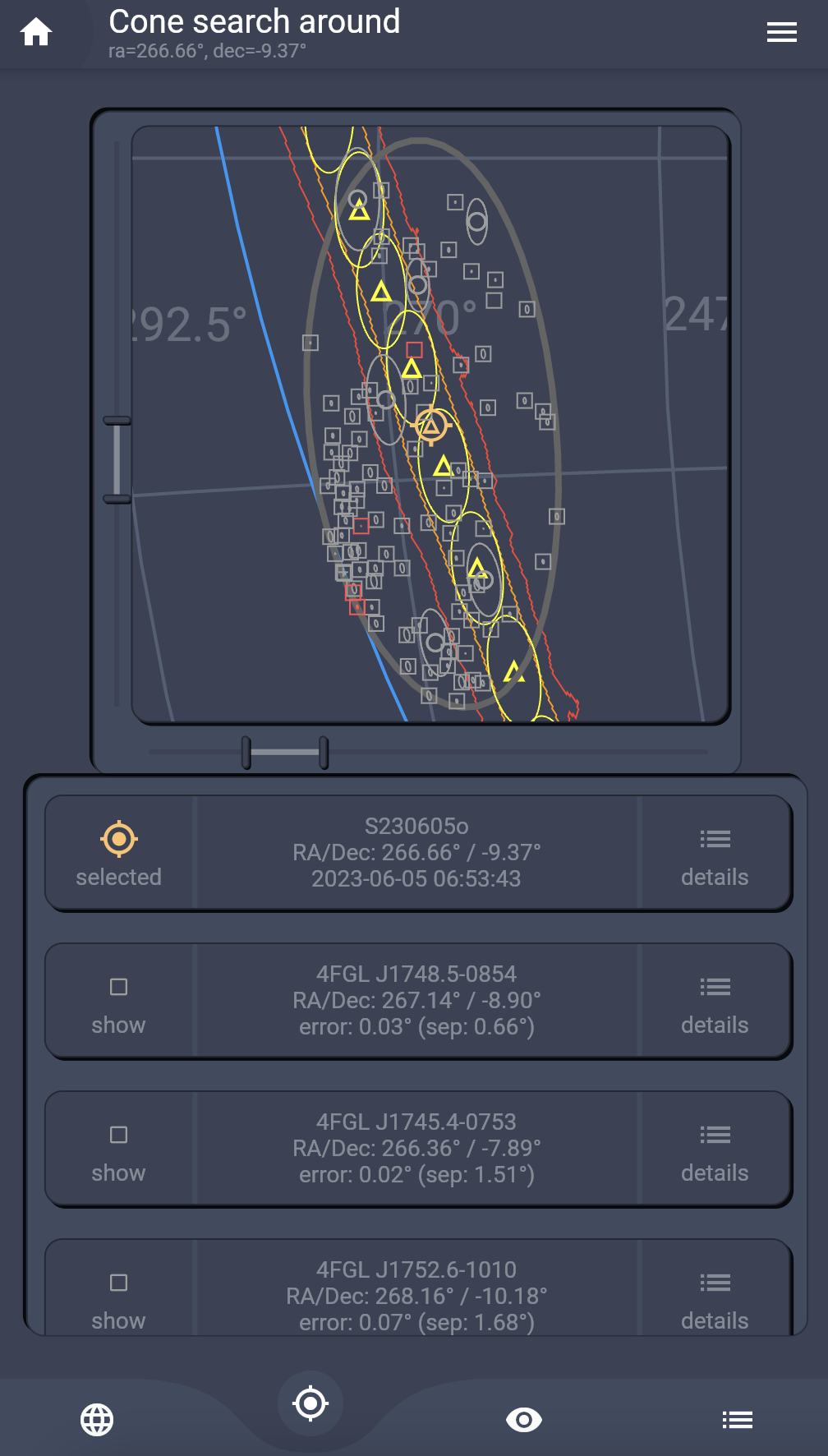} 
        \caption{Scheduling follow-upsof GWs with the mobile Astro-COLIBRI app.}\label{fig:tilepy}
    \end{minipage} \hfill
    \begin{minipage}{0.67\textwidth}
        \centering
        \includegraphics[width=0.9\textwidth]{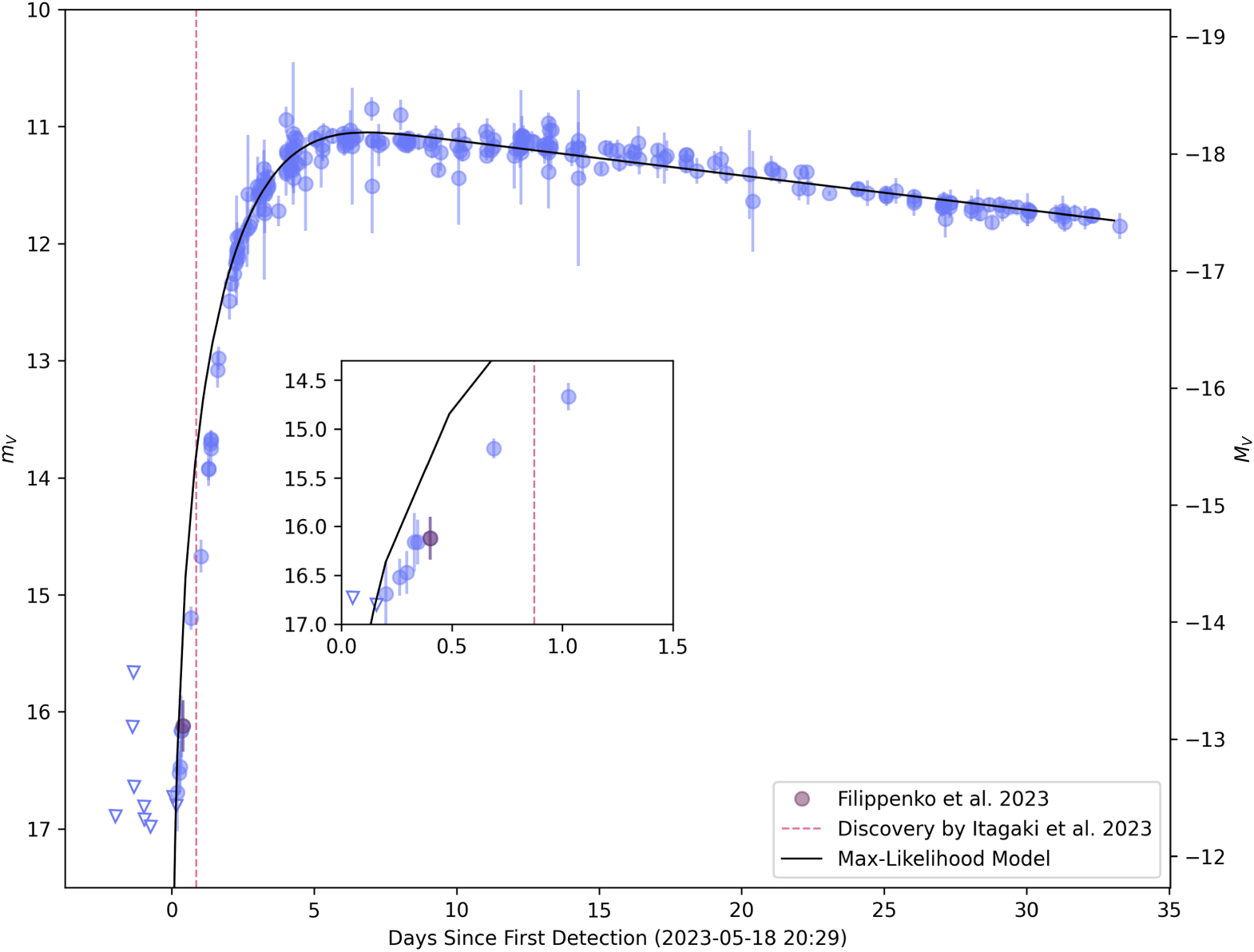} 
        \caption{Lightcurve of SN2023ixf obtained by the Unistellar Network of citizen scientists. From~\citep{Sgro_2023}.}\label{fig:sn2023ixf}
    \end{minipage}
\end{figure}

Furthermore, the {\it tilepy} API has been integrated into the Astro-COLIBRI graphical interfaces. A typical use case can be described as: if matching the various user-defined filter criteria, Astro-COLIBRI users are notified upon the detection of a new GW event in real-time. They can verify all details and parameters of the GW events directly in the app. The localization of the event is displayed via two contours, representing 50 and 90\% containments. For each Gw event and each user-selected follow-up observatory, an optimized follow-up schedule can then be obtained directly on a smartphone or via the web interface. The derived observation positions are then automatically overlayed on the skymap which allows to check for possible related detections of transient events (e.g. a GRB, a high-energy neutrino, etc.). An example of this display is shown in Fig.~\ref{fig:tilepy}.

\subsection{Citizen science: SN 2023ixf}
Although Astro-COLIBRI has been developed by and for professional astronomers, it has seen rapid and wide adoption in the amateur astronomy community. Àlthough holding increasingly sophisticated equipements, the threshold of accessing the streams of transient alert information through the various brokers and distribution networks like GCN, TNS, AMON, VSNet, AAVSO, etc. was too high for meaningful participations in the studies of transient phenomena. The ease of access provided by the Astro-COLIBRI interfaces coupled with dedicated information campaigns (e.g. on social media) and the organisation of workshops dedicated to amateur astronomers has significantly changed this situation. Citizen scientists are now providing important observations of supernovae and other time-domain phenomena (cf. \url{https://astro-colibri.science/usecases}).

A recent example is the observations of SN 2023ixf, a type-II supernova in the nearby M101 galaxy. Amateur observations of the galaxy allowed not only to detect the supernova very early in it evolution~\citep{2023TNSTR1158....1I}, they also allowed to pin down the explosion time to a window of only ~1h~\citep{2023TNSAN.133....1Y}. This in turned will allow to substantially increase the sensitivity of searches in neutrino detectors like SuperK~\citep{2023ATel16070....1N}. The bright supernova was also accessible with small telescopes such as the 11.4-cm aperture Unistellar telescopes~\citep{2020AcAau.166...23M}. The resulting light-curve has a 3.3h average sampling time over 35 days and is shown in Fig.~\ref{fig:sn2023ixf}~\citep{Sgro_2023}.

\section{Conclusion}

In this contribution, we have presented Astro-COLIBRI, an advanced platform for real-time multi-messenger astrophysics. Astro-COLIBRI integrates real-time multi-messenger observation tools into a comprehensive and user-friendly graphical interface, allowing astronomers to quickly acquire an overview of relevant information associated with transient events. The platform supports a wide range of astrophysical source classes and facilitates collaboration and data sharing among the astronomy community.

We have outlined the main features of Astro-COLIBRI, including its API endpoints and graphical interfaces. The API allows for the integration of Astro-COLIBRI's information into external environments and tools, catering to various use cases. The graphical interfaces, available on the web and as mobile applications, provide customizable settings and filter options for a user-friendly experience.

Through use case examples, we have demonstrated the power and effectiveness of Astro-COLIBRI. The platform has enabled searches for gamma-ray counterparts to high-energy neutrinos, facilitating rapid correlation analysis and triggering follow-up observations across the electromagnetic spectrum. It now also supports the scheduling of follow-up observations for gravitational wave events, optimizing observation positions for various follow-up observatories. Additionally, Astro-COLIBRI has seen widespread adoption in the amateur astronomy community, empowering citizen scientists to make important contributions to the study of transient phenomena.

In summary, Astro-COLIBRI offers a comprehensive and efficient solution for real-time multi-messenger astrophysics, enhancing the discovery potential and collaboration within the astronomy community. By integrating various data resources and providing intuitive interfaces, Astro-COLIBRI enables astronomers to better understand and explore the nature of transient astrophysical events.

The Astro-COLIBRI development team welcomes comments and feedback from the community to further improve the platform and can be contacted at \href{mailto:astro.colibri@gmail.com}{astro.colibri@gmail.com}.

\section{Acknowledgements}
The authors acknowledge the support of the French Agence Nationale de la Recherche (ANR) under reference ANR-22-CE31-0012. This work was also supported by the Programme National des Hautes Energies of CNRS/INSU with INP and IN2P3, co-funded by CEA and CNES and we acknowledge support by the European Union’s Horizon 2020 Programme under the AHEAD2020 project (grant agreement n. 871158). 

\bibliography{proceedings/Astro-COLIBRI_ICRC2023}{}
\bibliographystyle{aasjournal} %


%
%
%

\end{document}